\documentclass[journal]{vgtc}   
\ifpdf%                                % if we use pdflatex
  \pdfoutput=1\relax                   % create PDFs from pdfLaTeX
  \usepackage{graphicx}                % allow us to embed graphics files
  \DeclareGraphicsExtensions{.pdf,.png,.jpg,.jpeg} % for pdflatex we expect .pdf, .png, or .jpg files
\else%                                 % else we use pure latex
  \usepackage{graphicx}                % allow us to embed graphics files
  \DeclareGraphicsExtensions{.eps}     % for pure latex we expect eps files
\fi%

\graphicspath{{figures/}{pictures/}{images/}{./}} % where to search for the images

\usepackage{microtype}                 % use micro-typography (slightly more compact, better to read)
\PassOptionsToPackage{warn}{textcomp}  % to address font issues with \textrightarrow
\usepackage{textcomp}                  % use better special symbols
\usepackage{mathptmx}                  % use matching math font
\usepackage{times}                     % we use Times as the main font
\usepackage{cite}                      % needed to automatically sort the references
\usepackage{tabu}                      % only used for the table example
\usepackage{booktabs}                  % only used for the table example
\usepackage{balance}       % to better equalize the last page
\usepackage{graphics}      % for EPS, load graphicx instead 
\usepackage[T1]{fontenc}   % for umlauts and other diaeresis
\usepackage{txfonts}
\usepackage{mathptmx}
\usepackage[pdflang={en-US},pdftex]{hyperref}
\usepackage{color}
\usepackage{xcolor}
\usepackage{booktabs}
\usepackage{textcomp}
\usepackage{microtype}        % Improved Tracking and Kerning
\usepackage{ccicons}          % Cite your images correctly!
\usepackage{todonotes}

\usepackage{soul}
\usepackage{balance}
\usepackage{enumitem}

\usepackage{multicol}
\usepackage{float}
\usepackage{color, colortbl}
\usepackage{csquotes} %Elsie added this

\definecolor{lgray}{gray}{0.9}
\definecolor{orange}{rgb}{1,0.5,0}
\graphicspath{ {figs/} }

\onlineid{0}
\vgtccategory{Research}
\vgtcpapertype{theory/model}
\title{Roboviz: A Game-Centered Project for\\ Information Visualization Education}
\author{Eytan Adar and Elsie Lee-Robbins}
\authorfooter{
\item
 Eytan Adar and Elsie Lee-Robbins are with the University of Michigan, School of Information. E-mail: \{eadar,elsielee\}@umich.edu
}

\shortauthortitle{Adar \& Lee-Robbins: Roboviz: A Game-Centered Project for Information Visualization Education}

\abstract{Due to their pedagogical advantages, large final projects in information visualization courses have become standard practice. Students take on a client--real or simulated--a dataset, and a vague set of goals to create a complete visualization or visual analytics product. Unfortunately, many projects suffer from ambiguous goals, over or under-constrained client expectations, and data constraints that have students spending their time on non-visualization problems (e.g., data cleaning). These are important skills, but are often secondary course objectives, and unforeseen problems can majorly hinder students. We created an alternative for our information visualization course: Roboviz, a real-time game for students to play by building a visualization-focused interface. By designing the game mechanics around four different data types, the project allows students to create a wide array of interactive visualizations. Student teams play against their classmates with the objective to collect the most (good) robots. The flexibility of the strategies encourages variability, a range of approaches, and solving wicked design constraints. We describe the construction of this game and report on student projects over two years. We further show how the game mechanics can be extended or adapted to other game-based projects.}


\vgtcinsertpkg

\begin{document}
%\finalstate



\maketitle

%-------------------------------------------------------------------------
\section{Introduction}
A hallmark of most modern information visualization courses is the final or course project. At many institutions these are intensive, group-based efforts that represent a significant portion of the course grade. Projects can be client--, student--, or instructor--driven. Ultimately, these are intended to challenge students to apply the theory and methods they learned to more realistic scenarios. Thus, the projects themselves serve as a learning experience that can both teach new skills and reify abstract content. Projects also serve as a learning assessment and enable an instructor to determine if high-level learning objectives have been met (e.g., an ability to construct a visualization system from end-to-end). 

Projects are common across many of the syllabii we observed (e.g., Berkeley's INFO247, UW's CSE512 and INFO474, Stanford's CS448B, Georgia Tech's CS 7450, UMD's CS734 Utah's CS6964, Harvard's CS171, UBC's CPSC 547, MIT's 6.894, Calgary's CPSC 683, and many others). However, just as visualization is a wicked design problem, so too is visualization pedagogy. Thus, while the course `project' is ubiquitous, instructors vary aspects of the projects in response to their learning objectives, philosophies (e.g., project realism), practical constraints (e.g., student skill level and course duration), and ultimately whether they serve an assessment function. Variations in the projects may include whether they are group or individual, the duration of the project, different milestones, differences in intermediate peer and faculty feedback, whether projects come from a theme or specific datasets, whether the projects have external clients, and so on. There are no inherently bad assignment designs. Most have evolved to fit the specific situation in which they are embedded. Our own course utilized a variant of this assignment type for a number of years (open-ended, potentially client-driven, group project).

A challenge for our course has been that open-ended projects can be problematic in that students end-up working on `visualization adjacent' problems (e.g., collecting or cleaning data, working with under-specified client goals, etc.). It is important to learn to deal with these problems as they often come up in real-world scenarios. However, through some mix of bad luck or choices students may not focus on learning the `right' things thus leading to poor learning and assessment outcomes in time-restricted situations.  That is, students spend excessive amounts of time on what the instructor considers adjacent problems rather than on good design or evaluation of what should be the \textit{main} focus of the course: the visualizations. Because of varying constraints in data and tasks, students had unequal opportunities to design and build encodings and interactions (e.g., some datasets may have a network component and others do not). Additionally, we have found that in many situations students are unable to adequately define the visualization tasks for their projects. Students highlight vague exploratory systems tasks that do not map to real problems and are un-assessable, both for the student and instructors~\cite{adarbanning}. With the move to online learning with COVID-19, we took the opportunity to both streamline our project assignment and address these concerns.

Specifically, we switched to Roboviz, a novel game-based project activity.  Student teams build interactive visualizations to use as a game interface to compete against other teams (the goal is to acquire robots on a mining planet). We designed the mechanics of the game to have (a) concrete objectives with various trade-offs, (b) a wide range of game strategies that encourage the creation of multiple encodings, (c) projects that can be evaluated both by the students (during game play) and by the instructional team. We provide a dynamic dataset that can either direct students to specific `optimal' designs or ensure that no optimal strategy/design exist. Our high-level task (win the game!) can be broken down to many well-motivated and assessable sub-tasks. Student grades are only very weakly tied to game performance (through extra-credit) and are judged through design and implementation criteria. Roboviz was designed for flexibility around learning goals. For example, we have offered a `communicative' variant akin to a sports visualization for game watchers rather than players~\cite{jin1997tennisviewer}. Roboviz was inspired by other `game' based capstone projects. These include MIT's famous 2.007, where students design a robot to play a competitive `game' against other teams in the course (e.g., to collect as many balls as possible). Similarly, MIT's Software Engineering course 6.170 for many years included a project where students designed an anti-chess game to play against other students (the object was to lose). 

In Roboviz, students are transported to an imagined planet and recruit robot miners to work for them. Robots have different levels of productivity, and the objective is to recruit the most positively productive robots by the end of the ten minute match. Over the course of the game, participants are ``fed'' pieces of data of multiple types at different stages of the match--time series, hierarchies, networks, multivariate, etc. Student teams are challenged to create a dashboard for live game play to compete against another team. An effective student dashboard enables certain ``analyst'' tasks, specifically enabling strategies around data collection or predicting certain outcomes.

In this paper, we expand on the specifics of Roboviz and reflect on its use over two semesters (over 21 teams and nearly 100 students). We find, for example, that teams generate a diverse set of data encodings. We also identify limitations introduced by the game mechanics on the types of projects created. We reflect on how the game can be modified to address this and other pedagogical objectives. We observe that student enjoyment of the project has seemingly improved with Robobiz (unlike past years, students did not need course staff to resolve issues with clients or team disputes). Finally, we reflect on the properties of Roboviz that can lead to alternative games. We hope that this approach will be used to create a wide array of game-based projects that can be used and shared across courses. A video introduction to Roboviz and student materials are available at \url{http://roboviz.games/viz21/}\footnote{Interested instructors are welcome to contact us for the back-end and generator code.}.

\section{Related Work}
Various efforts have identified the goals~\cite{domik2000we} and broad curriculum for information visualization courses~\cite{domik2015acm, hanrahan05}. Modern visualization courses have settled on a largely consistent set of learning objectives~\cite{domik2015acm, hearstvis, kerren2008teaching,kerren2007workshop, owen2013visualization, rushmeier2007revisiting}. An example includes: ``An understanding of key visualization techniques and theory \dots Exposure to a number of common data domains and corresponding analysis tasks \dots [and] Practical experience building and evaluating visualization systems$\ldots$''~\cite{cse512}. Courses will often cover theoretical aspects (e.g., perception, design, data), specific approaches (e.g., network visualization, text visualization, geospatial, uncertainty, etc.) and for practical practice, often feature some kind of project(s). The specific topics covered in any given course will naturally vary based on the student populations (scale and type, e.g.,~\cite{elmqvist2012leveraging}), prior knowledge, faculty interests, and course limits (e.g., how long the course is).

Though the topics are relatively consistent, instructors have introduced numerous innovations for active learning and projects. These range from ideation on post-it notes~\cite{schwab}, critique~\cite{kosara2007visualization}, spec-driven workshops~\cite{vizitcards}, and data physicalization~\cite{willett2016constructive}. Multiple Panels (e.g.,~\cite{hearstvis}) and Workshops (e.g., the Pedagogy of Data Visualization Workshops at VIS'16 and '17 and IEEE VIS Workshop on Data Vis Activities at VIS'20 and '21) as well as dedicated channels (e.g., the \#topic-teaching channel for the Data Visualization Society) have all offered innovative approaches to teaching information visualization. Despite this innovation, a mainstay of most University visualization courses (both graduate and undergraduate) involves some kind of final project (e.g.,~\cite{burch2020more}). Project-Based Learning (extensively summarized in~\cite{krajcik_blumenfeld_2005}) is an established educational technique which is used broadly in engineering courses~\cite{dutson1997review,froyd2012five}, including those on visualization. 

A full discussion of experiential learning and its various off-shoots (problem-based learning, project-based learning, constructivist learning, inquiry-based learning, etc.) are well outside the scope of this paper. However, we note that a key limitation of many of these pedagogical approaches is that when taken to the extreme they embrace a kind of `minimally guided instruction'~\cite{kirschner2010minimal}. Despite their popularity~\cite{froyd2012five}, such projects often overemphasize a focus on the methods and process of creating a disciplinary artifact (e.g., interface), but with insufficient engagement with the theories and principles of the discipline that are needed when learning the field. Thus, while open ended projects may reflect the nature of the practice of visualization, they may not provide the structure that leads to learning in novice students. This problem has been recognized in HCI-projects specifically~\cite{reimer2003teaching}. Open-ended projects can not guarantee that all students learn the same things in the process of building the project nor can realistically assess whether learning objectives are achieved. For example, a team that only has access to small temporal data may not encounter either the challenge or opportunity to use certain interaction or encoding approaches. A team that cannot get good engagement from their client may not be able to iterate on designs or evaluate their ideas.

Success with open projects often come down to luck. Was the team lucky enough to get a good, clean, rich dataset with good clients and a clear problem definition? Failures may become apparent way too late for some teams and instructors. The results are projects that are incomplete, too simplistic (i.e., do not demonstrate broad competencies), or ill-suited to any task--potentially because the client has none~\cite{adarbanning}. Many HCI and visualization assignment designs attempt to manage project risks. For example, some elect to change from student-- to faculty-selected projects where the instructor selects the team or project~\cite{domik2016data,benblog}. This improves the odds that students can produce projects. However, real data remains problematic in that it is often hyper-constrained and may represent a limited challenge to students and assessment opportunity for the instructor. Other instructors attempt to manage specific project risks by adopting techniques to enforce more rigid design processes~\cite{reimer2003teaching,10.1145/3356422.3356447}, attempting to increase motivation (e.g., through service-learning projects ~\cite{borkin2017visualization} or leveraging public competitions/awards~\cite{romero2015case}), or using simulated clients or users~\cite{10.1145/1140124.1140142}.

We view Roboviz as a tool to significantly reduce the risks of open-ended projects while still maintaining many of their features. The instructor can limit the range of visualization-adjacent tasks (defined as they wish), control the tasks, data, clients, users, etc. While this comes at some cost of realism, it also provides fine grained control over learning and assessment. Our use of Roboviz is motivated by our goals of: (1) introducing concrete tasks~\cite{adarbanning}; (2) offering wicked design challenges; (3) pushing students to focus on visualization rather than related tasks; and (4) supporting the evaluation of student work. This last is often under-appreciated in assignment design. Many information visualization projects, both in and out of the classroom, are difficult to assess~\cite{plaisant04}. Our belief is that student projects should be easily assessable both by the students and course staff. 

Roboviz shares in the idea of using simulated data for projects~\cite{bertini}. However, Roboviz goes a step further by simulating not only the data, but also the tasks. In total, our assignment mechanics are not only designed to encourage variability in design, but also serve to assess overall student learning in the course: everything from implementation to theory.

\subsection{Competitive Projects in Engineering}
It is important to note that while Roboviz is a game, our course is not ``gamified'' in the traditional sense~\cite{Aguilar,huynh}. While utilizing game mechanics in the design of a course or around specific lessons is potentially an effective approach, this is orthogonal to our work. Rather, Roboviz was designed along the lines of competitive engineering capstone projects (e.g., MIT's 2.007 and 6.170 courses). The idea of \textit{creating} a game as a project is certainly not new. Many first projects in computer science involve coding up a game. These are viewed as fun for students and have well defined rules, making implementation achievable and assessable. \textit{Competitive} projects have been found to be effective for engineering education~\cite{chatzis, EGUCHI2016692, verner2000fire} and have evolved into global phenomena (e.g., the FIRST robotics competition). Experiences with engineering projects where competition was an option demonstrate more successful projects among those who chose to compete~\cite{jamieson2011early}. The use of Kaggle competitions in Machine Learning or Data Mining courses have a similar flavor. However, Kaggle competitions are often motivated by real world datasets. A related approach in visualization education is the use of VAST competition datasets in education~\cite{whiting2009vast}. There are other datasets that have been similarly used in competitions, including by the visualization community~\cite{4359491}. These competitions are focused (e.g., phylogenetic data) and narrowly constructed. The results are often novel visualization types, with a focus on a single encoding. Unlike these efforts, Roboviz data is purposefully synthetic. It provides both the opportunity and motivation to build many different kinds of visualizations across many different data types and tasks.

We note one related project idea devised at the Blekinge Institute of Technology in the context of a broader game-development curriculum~\cite{10.5555/3059068.3059070}. The particular visualization course was entirely connected to game development with a final project that involved visualizing game data. However, these leveraged existing games rather than a game specifically created for the course.

\subsection{Game and Sports Visualizations}
An additional appeal of a game-driven project is that students have significant examples and literature on which to draw. The visualization literature for games (e.g.,~\cite{agarwal2020bombalytics, bowman2012toward, wallner2013visualization}),  sports (e.g., ~\cite{du2021survey, jin1997tennisviewer,perin2018state}), and the combination (e.g.,~\cite{charleer2018real}) offer interesting starting points. However, we purposefully avoided the use of any specific existing game as inspiration for Roboviz. In part, this was a natural outcome of our design goals (varying strategies, data types, etc.). However, this also ensure that students aren't able to simply copy existing solutions.

\section{Course Context}\label{sec:context}
We briefly describe the context in which our course is run. While it is similar to other courses, there are a number of pedagogical and practical factors that impact our student's approach to the Roboviz game.
Our course is a medium-sized (50-80 students), graduate-level information visualization course (SI649). Programming experience is required, but this may range from two semesters to many years (e.g., those with computer science undergrad degrees). 

Our semester runs over 14-15 weeks. The first 7 weeks of the course are focused on `first-principles': introduction to visualization, data models, perception, design, communication, evaluation, and interactivity. The remaining weeks target specific data types: multidimensional, temporal, hierarchical, network, geovisualization, and text. These topics align with the data types that are used for different strategies within the game project. A complete week-by-week learning objectives document is available as supplemental information.

The course is taught in a flipped-classroom style. Students watch videos ranging from 45 minutes to an hour before attending a `lecture' session. Students annotate both the flipped videos and required readings. The 1.5 hour lecture session is taught through an active learning style. Students are put into randomly-assigned mini-groups to discuss questions around the week's topic before returning for a full-class discussion. A second 1.5 hour session during the week is focused on `labs'-- either programming for the first 5-6 weeks of the course (Tableau, Altair, and Streamlit in the 2020 \& 2021 iterations), or design work through the VizItCards format~\cite{vizitcards}. By the end of the lab sessions students could build and deploy a range of interactive visualizations using Altair. We emphasize these features as we explicitly create many opportunities for students to meet each other and work in different groups before committing to a final project team. Additionally, the students develop significant expertise in group design.

A mid-semester individual project is focused on a communicative visualization exercise. In Fall 2020, the project was based on an article explaining why Norway does well in the Winter Olympics. The Fall 2021 variant was around an article about quality of life in a rural US county relative to the rest of the country. 

The final project in the course has always been designed to allow students to apply all their learning to the construction of a new visualization system. With Roboviz we set our learning objectives as having students: 1) apply good design practice to generate and contrast design alternatives; 2) document and justify design decisions based on visualization theory; 3) implement functional and usable visualization systems; 4) select and utilize appropriate encodings for a range of data types and design tasks; 5) implement appropriate interaction techniques in visualizations; 6) apply good design to produce aesthetic visualizations. Note that our emphasis is on visualization concepts and implementation. These objectives are mirrored by the rubric (see Section~\ref{sec:rubric} and supplement). Ideally, a good Roboviz project should not only demonstrate the achievement of the learning goals above but also many of the other learning objectives of the course. For example, they will be able to determine if a small multiples display is warranted and what interaction techniques (e.g., linked scrubbing, highlighting, etc.) are most suitable given the design goals.

\section{The Roboviz Game}
The Roboviz game was designed to encourage students to build many different visualizations that can be used to either \textit{play} or \textit{describe} a game. Teams must figure out a game play strategy based on the game mechanics and build a visualization dashboard. Games are simultaneously played by two teams or one team against a simulated opponent. The complete description of the game and rules is available at \url{http://roboviz.games/viz21/} and in our supplement.

\begin{figure}[hbtp!]
    \centering
    \includegraphics[width=\columnwidth]{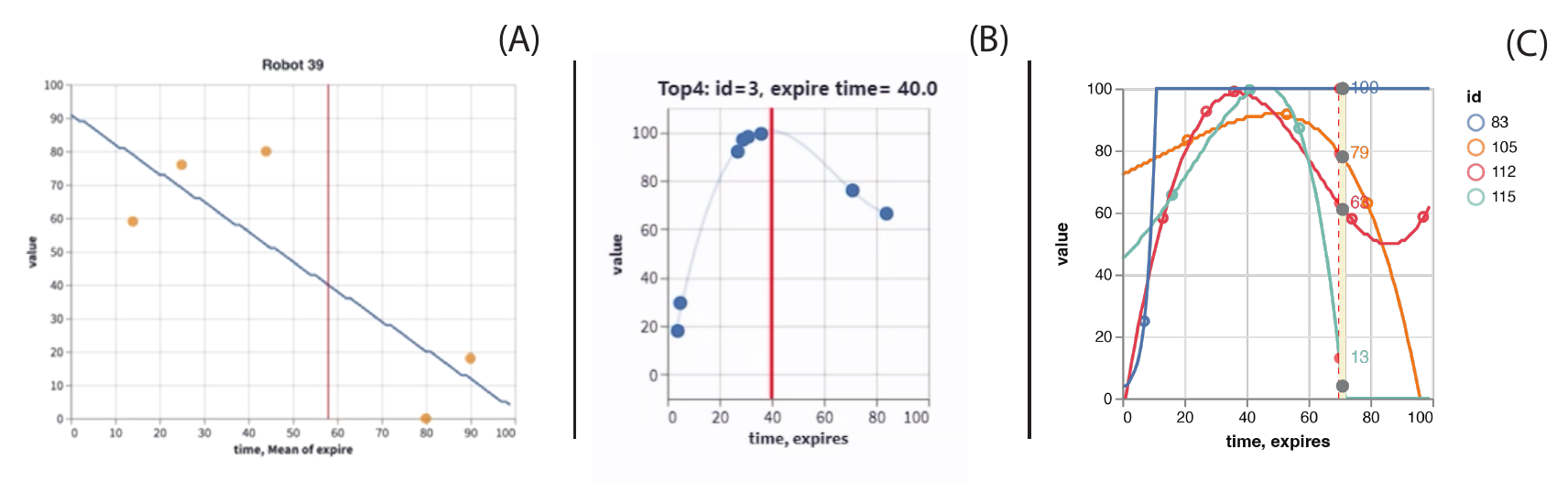}
    
    \caption{Screenshots for three versions of the scatterplot + line chart visualization for the time series data. Chart A is a basic scatterplot and line chart, with a straight line of best fit. Chart B uses a polynomial regression to smooth out the lines and better predict what the value might be. Chart C includes data from the robot's family tree to increase the predictive power of the line by having more data. 
    }
    \label{fig:timeseries}
    % SVGs to replace A and B?
\end{figure}

The high level mechanics of the game are to collect as many ``points'' (\textit{productive} robots, in our case) within a 10 minute match. At the simplest level, a team will guess a value that the robot is thinking of before the expiration (sometime \textit{within} the 10 minute match). A more accurate guess of the robot's random number will usually get you the robot. However, not all robots are productive, meaning that some will lower a team's overall score. To add complexity, some unproductive robots are valuable for other reasons, leading to multiple strategies on which robots to try to collect. Additionally, teams can ask for additional streams of data that can help them decide on which robots to try to collect and what numbers to guess. As we describe below, teams are gradually fed a little bit of this information (e.g., some timeseries data, some multivariate data) at each time step of the game. This temporal structure was used to better reflect many analytical scenarios. That is, decisions and analyses are often temporally evolving. Even in static datasets, one analytical choice may lead to the next.

As is hopefully apparent, there are \textit{many} viable strategies. The game was designed that even with a full team playing, it would be difficult to simultaneously track everything for an optimal strategy or build a visualization for every strategy and case. Teams must design their visualizations to support \textit{specific} strategies and are thus forced to tackle our first key learning objective--applying good design practice to generate and contrast alternatives. This is verified by evaluating their design process documentation.

The robot information comes in a variety of formats and correspond to different strategic mechanics: temporal, multivariate/multidimensional, network, and hierarchical. Each piece of information corresponds to one strategy. Roboviz is designed in a way that no particular strategy can dominate over the other. Teams must use at least two types of data--and corresponding visualizations to effectively play--as well as adding interactive elements to support their use. The game mechanics encourage this diversity of visualization and interactivity as it is impossible to play effectively without these elements. By varying the way the underlying randomization happens, we can also ensure that the combination of strategies needs to be different \textit{between} matches. Thus, to be super effective, a team should try to build interactive visualizations to support all the strategies.

The benefit of this mechanic is that it incentivizes teams to build visualizations corresponding to the many types covered in class. The goal for each team's `system' is understood, but teams can be highly creative in building the game playing dashboard. The complete rules of the Roboviz game and a video walkthrough are available at \url{http://roboviz.games/viz21/}. Here, we briefly describe key features of the game mechanics, and implementation details.

\subsection{Basic Rules}
Teams are told that they have landed on a Planet X421ZZ and have 10 minutes to recruit 100 robots. The robot ``miners''--which collect planetary resources--have evolved from those previously abandoned on the planet. Some robots are productive, some not. Thus, not all robots are worth trying to recruit, and unproductive robots will count against a team's final score. Game time units, called XTUs (Planet X Time Units), are six seconds. This gives us 100 XTUs per match, for a total of 10 minutes. Teams are explicitly told that they cannot automate any part of the game play. The team, or player, must make strategic decisions by studying their visualizations. Teams are given two datasets at the start of the game (the social network and the family tree) and gradually receive data points (temporal `friendship game' data and quantitative/categorical `parts data') throughout the game (see Section~\ref{sec:hacker}).

\subsection{Robot Friendship Game}
The main mechanism for recruiting robots is through the \textit{Robot Friendship Game}. 
Students are told that the robots have a pseudo-random number generator, and that they will join their team if they can guess closest to the `number they are thinking of.' Each robot has a random `expiration' time for which the team needs to guess by (see the horizontal red line in Figure~\ref{fig:timeseries} (B) as an example). 
Internally, for each robot, we generate a pseudo-random time series (a $4^{th}$ order polynomial with random coefficients in the current instance) for all 100 time units for the game (see the thin blue line in Figure~\ref{fig:timeseries} (B) as an example). The teams have a number of ways to get information about values along each robot's time series that allow a team to make an educated guess about the value at expiration date.

For example, Pushwalker Botson (robot ID 87--all robots have a name and ID), would like a guess by 60 XTUs after the game starts (360 seconds in). The true answer for that time point is 92. If team $A$ can guess or predict this value more accurately than team $B$, Pushwalker will join team $A$. Because new information is constantly being delivered up to the deadline, guesses/bets can be updated up to that point. For strategic reasons, it is also possible to bet $-1$ to indicate that the team doesn't want the robot. More specifically: (1) if both teams guess -1, the robot powers down and no one gets the robot, (2) if your guess is closest to the correct answer and the other team is not within 10 of the correct answer, you will get the robot, (3) if both teams are within 10 of the correct answer, the robot will decide who to join by using the social network strategy (described below).

\subsection{The Social Network}
The students are told that the robots have evolved their own social network which they use to help decide which team to join. If both teams are close to a robot’s true answer in the friendship game (or there’s a tie), the robot will look to their social network to help them decide. In the current implementation, the social network is generated as a powerlaw graph (specifically~\cite{PMID:11863587}). In the basic implementation, each of the 100 robots are placed randomly in this graph.  The entire network is provided to teams at the start of a match. 

\begin{figure*}[t]
    \includegraphics[width=\textwidth]{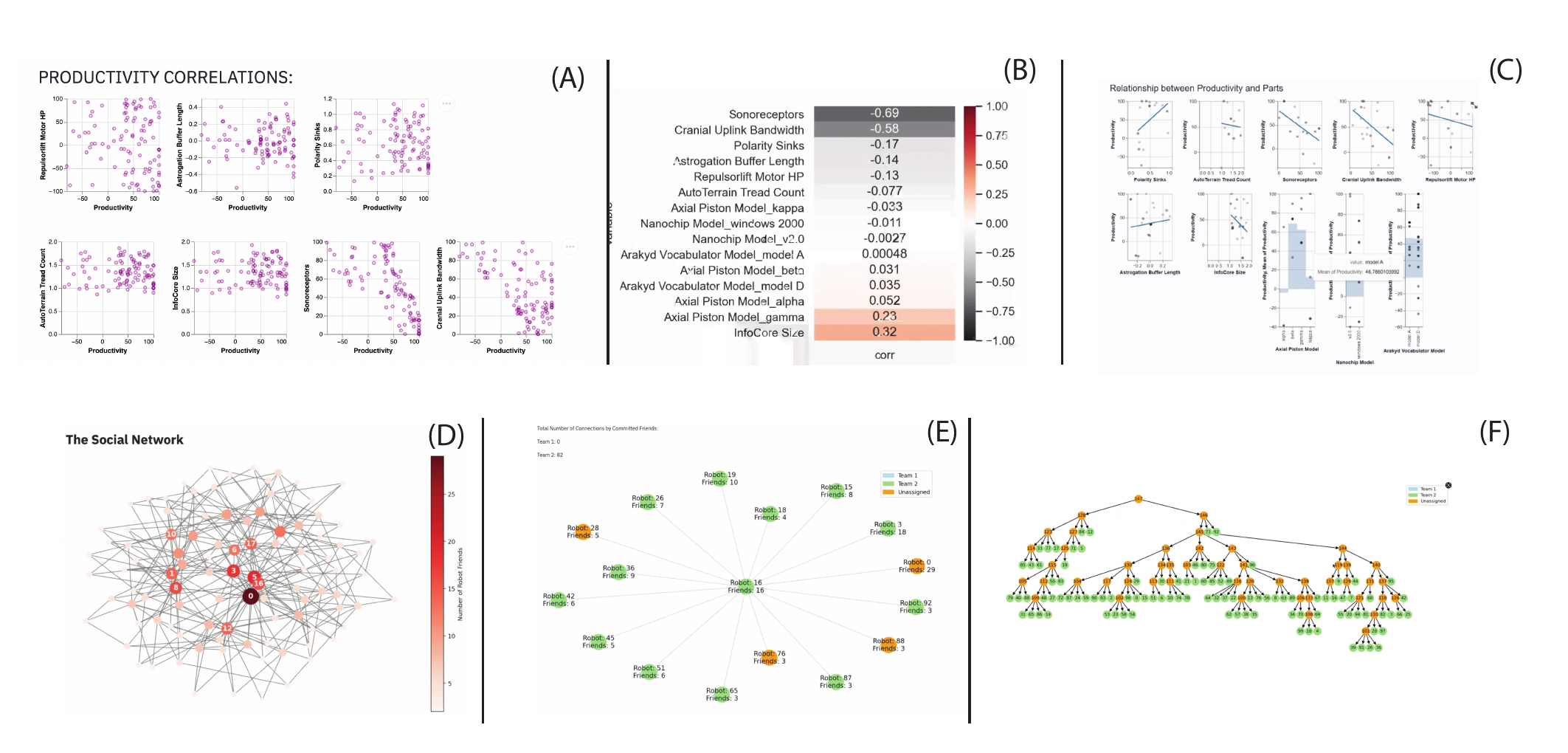}
    
    \caption{Sample visualizations produced by students teams. The top three (A, B, C) were different ways to visualize the correlations between specific parts and productivity (treating these as independent). The bottom are example visualizations of the network and hierarchical data from one team's dashboard (at the end of the match). The team would study a visualization of the social network with robots color encoded based on in-degree (D). When they noticed a robot about to expire they would pull up a detailed view (E) to study its neighbors. Visualization F is a visualization of the robot family tree for the match (this was from a simulated game so all robots were claimed by one team). 
    }
    \label{fig:networkhier}
\end{figure*}

When deciding between two close bets, a robot will consider those immediate neighbors that have already joined one of the two teams. They will take the weighted average of the neighbor based on degree (a more ``popular'' robot will have more weight). This additional mechanic encourages visualization of the network data and consideration of which robots to prioritize. Two example visualizations of this data from an actual interface are captured in Figure~\ref{fig:networkhier} (D,E).

\subsection{Robot Productivity}
In our game, robots are not equally productive and productivity is a function of their parts. Students are told that robots consist of 10 parts/features that they should care about. The first 7 are quantitative features (e.g., 7.3 or -92)  and the last 3 are nominal/categorical (e.g., Alpha, Beta, Gamma). Some robots may even have negative productivity. Recruiting a bad robot will hurt the team's overall performance and counts against their score.  This creates an incentive to determine both the relationship between productivity and a part (do they help or hurt) and what parts a robot has.

Specifically, we pre-generate a linear combination of the various variables to produce the final productivity (ranging from -100 to 100).  At random, some variables are determined to be highly correlated. Strategically, teams would like to learn which parts are predictive of positive or negative productivity robots.  In the implementation of our game so far, we have set the average productivity of all the robots to be greater than 0. Because of this, there is some advantage to collecting as many robots as possible.

\subsection{The Family Tree}
As a final level of data complexity we also introduce a hierarchical data structure. Teams are told that in mimicking human societies, one robot, long ago, decided that it wanted to ``evolve.'' Over time, old robots built newer ones and a family tree emerged (teams are given this data at the start). Internally, this tree is produced by randomly grouping sets of 2-4 robots and assigning them a `parent' (this is a fake internal node that isn't one of the 100 robots). Robots assigned to a group are removed from consideration and replaced by their parent node. The cycle continues until there only one `ancestor' robot remains. This process ensures that the 100 robots in play are only in the leaf nodes. A visualization of one such structure is in Figure~\ref{fig:networkhier}(F)). The hierarchical structure connects back to the robot guessing game. Related robots will have similar time series to each other. By knowing where robots sit in this family tree, it's possible to make a better guess if we know things about `sibling' or `cousin' robots. It is also possible to use the hierarchical visualization to target certain well-populated `sub-trees' for additional data. Strategically, a visualization can help recognize these robots.

\subsection{The Hacker}
\label{sec:hacker}
Some dataset pieces are provided to the teams at the start of the match: the robot names and ids, the social network, the family tree, and each robot's deadline for the guessing game. Once a robot has `expired,' teams are also given its productivity value. All other information is gradually provided in a pseudo-random fashion by the `hacker.' 

Students are told that the hacker is an agent that has broken into the robots' databases and is slowly leaking information to the team (each team has their own hacker). Each time unit (6 seconds), the hacker can provide values from the robot's random number generator (e.g., the light sensor for robot 5 at time 40 was 9) and part values (e.g., robot 54's Astrogation Buffer Length is 23). The hacker delivers 10 data points at each step (5 each from the time series and parts data). Of course, this is simply sampling from the various pre-generated data frames. Teams can ask the hacker for information about specific robots or specific parts. This allows them to build models strategically (e.g., productivity predictions or friendship game expectations for unexpired robots, or for building productivity models for all robots). The server will bias the produced `hacked' data during game play towards those robots or parts. Teams may continuously update their requests, but only get new information every 6 seconds.

\subsection{Match Construction}
All the data for a match is pre-generated. A script produces all the necessary match files which are then loaded to the server for a given match. No two matches will be the same. To develop their team visualizations, students are provided with a few practice match files, a simulated competitor (including code), and a simple server to handle the match. The simulated player has all of the match data files, so it is omniscient. However, it makes random guesses for the friendship game within some range around the true answer. In theory, it should be easy to beat but presents a useful starting opponent for teams to test their ideas and develop a game strategy. Teams can extend the simulated player to be more challenging (i.e., it can be less error prone).

During a match, the game play is recorded by the game server. This includes all data requested by each team, the data provided by the server, the robot `joining' behavior (and reasons why). This log allows us to study (or visualize) completed matches in full detail.

\subsection{Client APIs and Simulators}
In addition to the server and simulated player, students are also provided with client APIs written in Python and in Javascript. Additionally, we provide documentation for the JSON-based API for those teams not working with one of those two systems. The client APIs allow teams to log onto the match, collect `public' data (e.g., the network, hierarchies, current robot commitments, etc.). Additionally, the client allows teams to issue requests to the hacker or bids for the robots. We provide students with examples written in Python and Javascript that demonstrate how different interfaces can be built with the APIs. All students in the course are experienced with Altair/Vega-Lite and with Streamlit. As we discuss below, most of the interfaces are built using these tools. We also provided reference to more specialized visualization libraries that were not previously covered (e.g., nx\_altair for network visualization). We make no particular constraints here on how the systems are to be implemented. To get students started, we have created a video walkthrough for both the game and the use of the API to generate simple visualizations. We found this useful as some students are less familiar with API-based programming. The video is available at \url{https://www.youtube.com/watch?v=xNdjV2cWj7U}.

\section{Course Results}

We have had the opportunity to run the complete Roboviz project during the Fall'20 and Fall'21 semesters. Due to COVID-19 restrictions, the course was entirely online in Fall'20 and in a hybrid format in Fall'21. However, in both semesters we ran the Roboviz competitions online. 

\subsection{Assignment Details}
Students were given four weeks at the end of the semester to develop their dashboards and strategies before the final product was due. We suggested groups of 3-5 students as this allowed each member to tackle a different encoding type, was consistent with our other courses, and is generally viewed as best practice (e.g.,~\cite{laughlin2006groups,benblog}). As an accommodation for COVID-remote learning we allowed for individuals or two-member teams to participate (we had one individual participant in Fall'20 and a two-person team in Fall'21). Students were prompted to find a good mix of skills when selecting teams and most students had interacted in earlier small-group activities in our flipped environment. In the first implementation of this project, teams were allowed to either focus on a communicative role (e.g., presenting post-game analysis) or a player (i.e., analyst) interface. Of the 11 Fall'20 teams, six chose to build the player tool. In the second iteration in Fall'21, we discontinued the communicative option and all ten teams developed game play interfaces. This was done to encourage every student to experience one communicative project (achieved through the mid-semester individual activity) and one more focused on analysis/exploration. Courses that do not have this balance could certainly retain both options for Roboviz. 

Teams could build their system using whatever tool they preferred. The bulk utilized Streamlit, two used a standard Jupyter Notebook, and two teams used Tableau.  
Students largely utilized Altair as the basis for visualizations (followed by matplotlib, Seaborn, Pyvis, and a smattering of other libraries). We asked all students to run the VizItCard workshop~\cite{vizitcards} to rapidly iterate over a solution space. Students had experience with this format from past labs. Teams who were focused on a communicative project were instructed to craft a set of communicative learning objectives~\cite{lovis}. 

Because of the hybrid teaching environment, some teams were distributed internationally. To support these students, we offered the option to either play against other teams during large sessions or to make an appointment with the course staff and play against a simulated player. 
By Fall'21, most students remained on campus and 8 out of 10 chose to play in a bracket-style tournament. We scheduled two sessions, each about an hour and a half long, with three matches in each session.  

For students who participated in the communicative variant (F'20), we provided trace logs of a few simulated matches (as no games had been run at the time). All groups were asked to create a video presentation of 10 minutes or less describing their project. They were instructed to describe their goals, early designs, ultimate decisions and offer a walk through of their implementation. Students submitted this video report as well as all code.  We used the video report format in part to reduce student burden given COVID-related changes. However, we have found that overall this is a positive change and worth retaining. Students demonstrate their project and describe design decisions in sufficient detail for course staff to understand and assess the work.

Students had a number of approaches to both development and game play. Tournaments and matches were conducted over Zoom. Each team was given a breakout room to coordinate through. The `main' room displayed a running score card. Observers were allowed to enter the game rooms to watch teams. One team physically co-located to play, but most teams played remotely. The bulk of teams utilized multiple people to play for their team simultaneously. Given the significant stream of data---far more than one person could likely handle, even with the best interface---this was strategically advantageous. We observed significant variation across teams in collaborative designs and play. Roles varied across teams where some had multiple people focusing on different robots whereas others had different tasks (e.g., one inputting guesses or hacker requests with others analyzing data). Teams collaborated through digital coordination (e.g., chat) but also simply speaking to each other. Some shared one visualization in the Zoom window, but most utilized local dashboards for each person. Retrospectively, it's not clear to us that any one strategy was dominant. However, we were encouraged by all the variations in play.

\subsection{Visualizations created}
\begin{figure*}[hbtp!]
    \includegraphics[width=\linewidth]{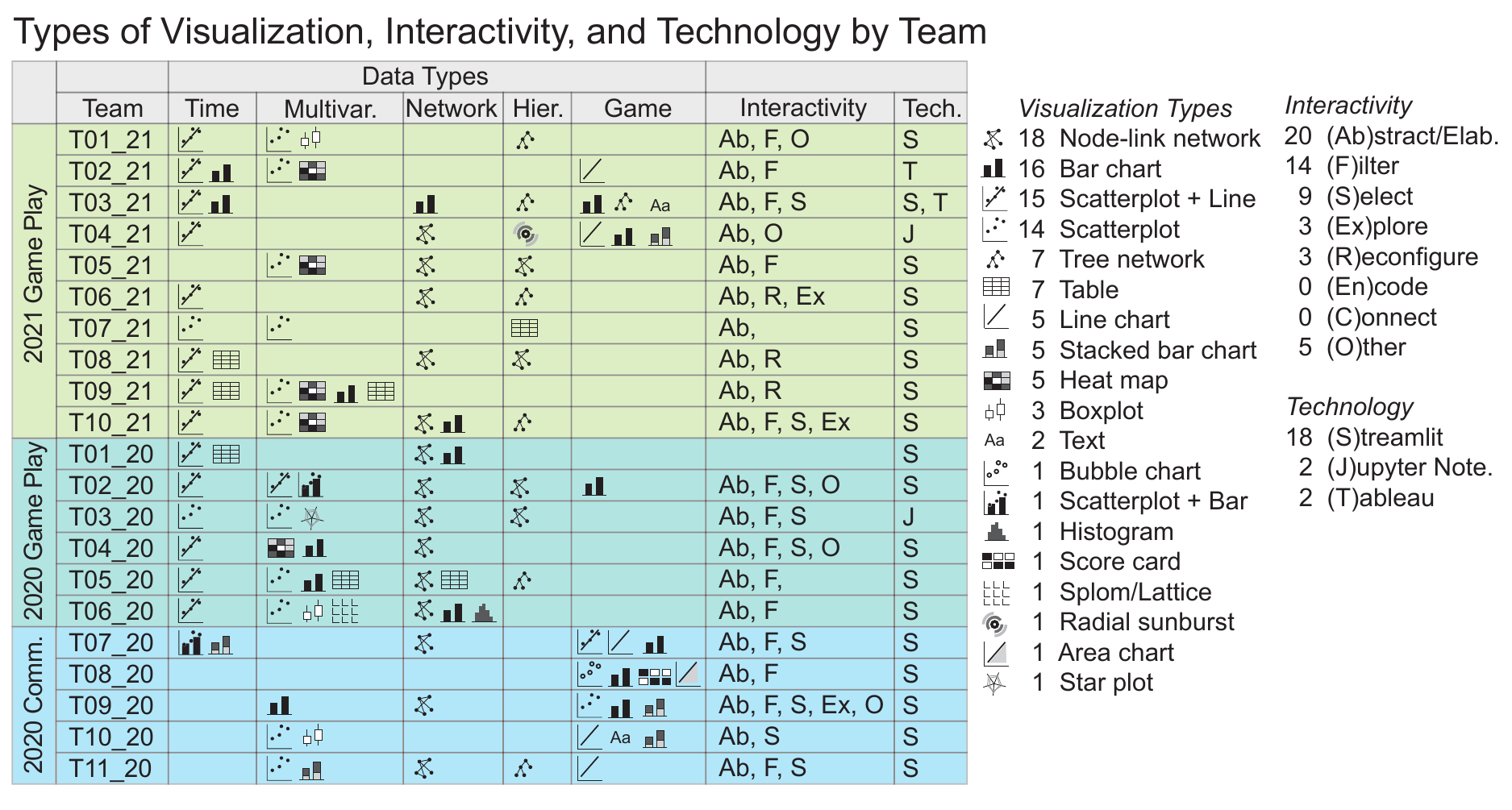}
    
    \caption{The different visualization, interaction types, and underlying technologies observed across all participating teams. 
    }
    \label{fig:types}
\end{figure*}

With respect to design, the Roboviz game involves four data sets (time series, multivariate, network, and hierarchy). Figure~\ref{fig:types} summarizes all the observed visualizations, interactivity elements and technologies used by our teams. We required groups to choose at least two datasets to visualize. We found that 12 (out of 21) groups chose to visualize three datasets, 4 groups chose to visualize two datasets, and 4 groups chose to visualize all four datasets. In addition to the datasets about the robots, 4 analytic teams also chose to visualize a game tracker, which included metadata about who was winning the game, their productivity scores, or which teams won which robots. For obvious reasons, all 5 communication-focused teams visualized a version of a game tracker.

\textbf{Time series} Fifteen (out of 16) analytic teams visualized the time series data. This is unsurprising because the main task was to figure out the robot's random number for the friendship game --- and the only way to do this was to look at the time series data. All teams that visualized this data used a scatterplot, with most (13) of those teams also using a line chart overlaid onto the scatterplot (see Figure~\ref{fig:timeseries} for examples). Although there was much uniformity with the types of charts for this dataset, the details of the charts, and thus the utility of the chart, varied by team. 
Some common design features of these charts were a polynomial line of best fit and a vertical line indicating the robot's expiration time. Some of the best versions of these graphs used small multiples to show several robots at once, automatically filtered the robots to only show the next 5 robots that are `expiring,' and annotated the best guess for the random number on the graph. Some of the least effective versions only had scatterplots or required manual (de)selection of robots from a drop down list.

\textbf{Network}. Fifteen teams visualized the robots' social network. Teams that recruited popular robots could be more likely to also recruit their friends as well. The most common visualization type for this data set was a node-link network diagram. These visualizations commonly used color, size, annotations, and/or tooltips to encode popularity. Teams also incorporated information on which team won the robot and the robot's expiration time to improve the usability of the visualization. Some teams chose to visualize the entire social network at once, while others only visualized a subset at a time. Some of the best versions of this graph were linked to other graphs, automatically highlighting the robot of interest to be found easily. Less effective versions were crowded, labeled poorly or not at all, and were difficult to navigate. 

However, some teams opted for another type of social network visualization, citing usability for their game play strategy. For example, Team T03\_21 created a bar chart of the 15 most popular robots (in the social network sense), sorted in descending order. In their video documentation, they explained, ``\textit{We translated the idea of doing a network graph into a bar graph of just showing who has the most relationships. Instead of having to parse through a network, we can just simply go after those with the highest network (popularity).}'' This choice and subsequent explanation reflect a deeper deliberation of how they were using the visualization and how their encoding choices could optimize their decisions. 

\textbf{Multivariate}. The multivariate data featuring productivity factors was the third most popular dataset to visualize, with 14 teams choosing this data set. Though not essential, productivity was a key factor in what makes a robot worth bringing onto your team. Most teams (12) chose to use a scatterplot, and most teams (11) used more than one visualization to show this data. The other common visualization types were heat maps, bar charts, and boxplots. A common implementation was for teams to create small multiples of scatterplots for the quantitative robot parts and a second type of visualization to show the productivity of the nominal robot parts (see Figure~\ref{fig:networkhier}(A) and (C)). Some ways that teams went above and beyond were to graph a line on top of the scatterplot showing the correlation. Some teams skipped the scatterplot entirely and just graphed the correlation values onto a heatmap (see Figure~\ref{fig:networkhier}(B)). Ineffective pitfalls were that some teams had the same $y$-axis for all of the robot parts, but some have ranges from 0 to 100 and others have ranges of 0 to 1, leaving the small ranges unreadable.

Although we asked teams not to automate their solutions, teams were still allowed to calculate different models for display. One unique team (T01\_21) used visualizations to estimate the relationship between different robot parts and productivity. Through an interactive interface, they fed that relationship back into their program and used it to approximate the equation from the game that decided productivity. They used this to predict the productivity of the robots that were not yet declared for. This was a very successful method and was one of the more creative approaches we have seen to this game.

\textbf{Hierarchical.} Twelve teams visualized the robots' family tree hierarchical data. Most of these visualizations were tree networks. 
In combination with the hierarchical data, some teams also encoded productivity or which team won the robot. Like the social network data, some teams only visualized a subset of the family tree, centered on a specific robot of interest. The most useful implementations of the family tree data was incorporating it with other datasets. For example, some of the time series scatterplot+line graphs automatically included the robot of interests' nearest family members as well (see Figure~\ref{fig:timeseries}(C)). The robots' random number generators were similar to their siblings and parents, so including that data give the teams more information to help them more accurately make a guess.  

\textbf{Game Tracker.} Four analytical teams and five communicative teams created visualizations to show game progress. A common representation was a line chart with different colored lines representing the productivity of both teams. Another way to represent the game score was either a bar chart or a stacked bar chart. As opposed to more elaborate setups, some teams simply had a time tracker at the top to keep them on schedule. Even for teams that did not have a dedicated game tracker, encoding which team was winning which robots in conjunction with some of the other datasets could also serve as a tracker. 

These visualizations provided real-time tracking of who was winning the game. This served both a functional role (seeing if a strategy was working) and made the game more fun to watch and play. For the communicative teams, game tracking was an essential element, giving an overview of how the game progressed over time.

\textbf{Interactivity.} The assignment specified that the students needed to include interactivity in their dashboard. They were evaluated on how well they implemented interactivity and if it improved the usability of their visualizations. We analyzed the interactivity that the teams used based on the categories in this paper~\cite{interaction2007}: Abstract/Elaborate, Select, Explore, Reconfigure, Encode, and Connect. 

The most common interactivity used was a hover tooltip (Abstract/Elaborate) --- this was used by every team (except for one team that did not use any interactivity). Often, the hover tooltip provided robot ids, more details, or specific data values on demand. The hover tooltip is also the easiest to add in Altair (and most other libraries). The second most common use of interactivity (14 teams) was to filter the data and show something conditionally. Teams used a variety of interaction methods to filter to see only a specific robot or a specific robot part. This was important, as they had to make decisions about one robot at a time, and there were 100 robots-- too many to visualize all at once. The third most common use of interactivity was to select and mark something as interesting. Teams used selection to highlight robots of interest to make them easier to see across several graphs. Infrequently used forms of interaction were Explore (some teams used a pan feature) or Reconfigure (changing facets, sorting tables). Two types of interactivity ((Re)encode and Connect) were not observed. 

\begin{figure*}[hbtp!]
    \centering
    \includegraphics[width=.8\textwidth]{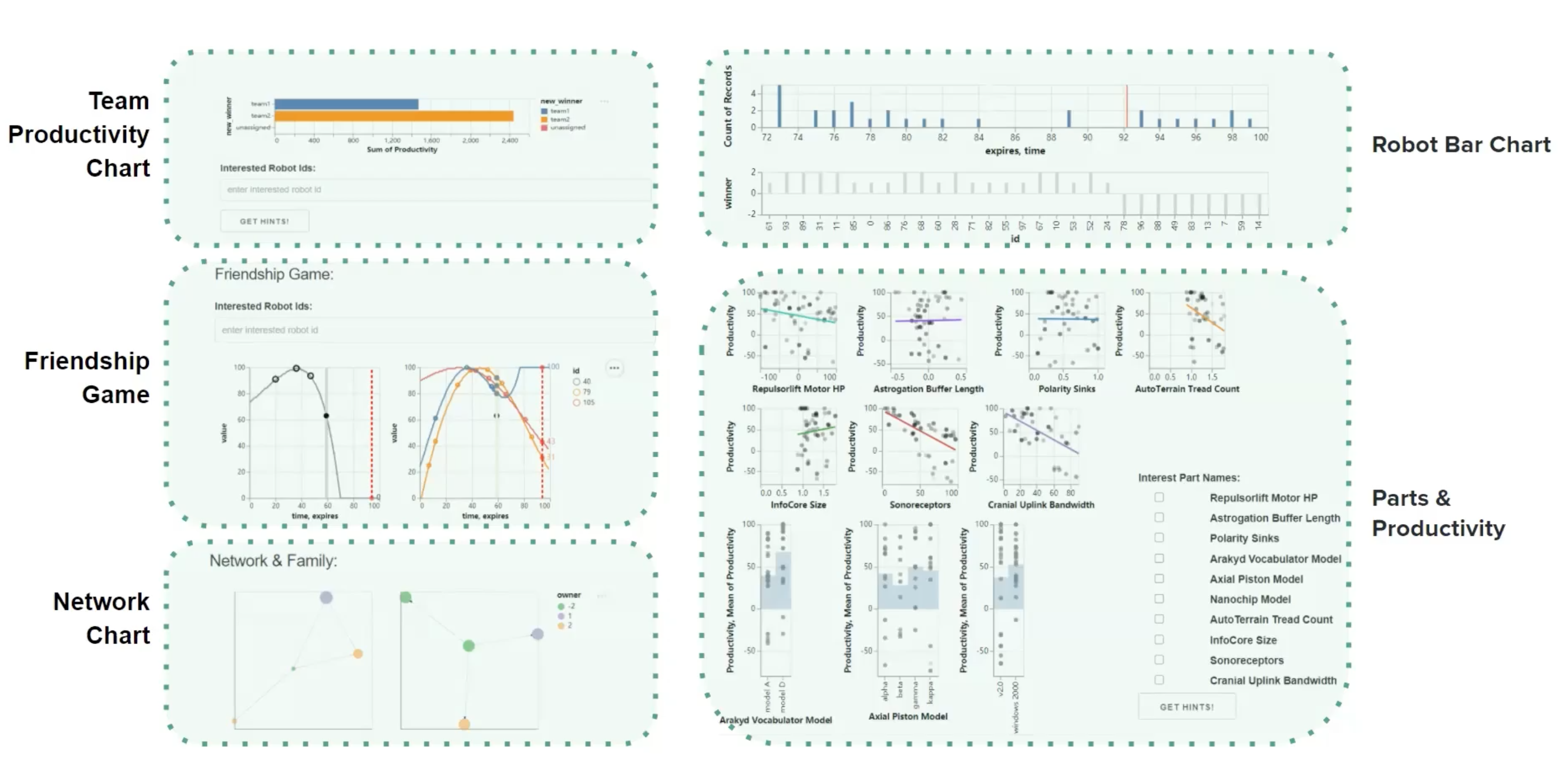}
    
    \caption{An example dashboard screen for one team. The annotations in the figure were created by the student team for their video presentation. In this case, the running dashboard was displayed on one screen and allowed one student to view and control game play. Team Productivity Chart: This was a game tracker bar chart that compared Team 1's score to Team 2's score. This also has interactivity for inputting robot ids for the hacker. Robot Bar Chart: This is also a game tracker which shows how many robots are expiring at what times and who has won those robots. Friendship Game: These are line charts + scatterplots with data from the hacker showing the robots' pseudo-random number generator time series. Network Chart: These node-link diagrams show the robot's closest family members and their friends. Parts \& Productivity: This section includes small multiples of scatterplots for the quantitative variables (top 7) and categorical variables (bottom 3). Both have summary lines or bars to indicate the correlation or averages over all the data. Interactive features include extensive tooltips on all the figures, dynamic lines on hover in the time series, and various search/filter features to focus on specific robots.
    }
    \label{fig:dashboard}
\end{figure*}

In addition to these types of interactivity, we also saw a few implementations of interactivity that did not fit into these categories. For example, the productivity prediction mentioned previously was a unique way to integrate interactivity into the system. Additionally, some teams had features in their dashboard to input information about what data they wanted from the hacker. These were interactive features, but they were specific to playing the game and not necessarily directly related to a visualization. Finally, two teams used a time slider to manually slide the game time elapsed. This worked well in the communicative dashboard, where this enhanced the post-game analysis to see the game in stages as it progressed. However, it resulted as mostly a superfluous addition to the analytical dashboard, as it did not help the team play more effectively. A complete, and somewhat sophisticated dashboard, is displayed in Figure~\ref{fig:dashboard}.

\subsection{Rubric}\label{sec:rubric}
We retained roughly the same grading rubric for the two semesters. We briefly summarize the one used in F'21. This rubric is clearly biased to our learning objectives and can be modified to emphasize other goals. Students were able to see the rubric through our learning CMS (Canvas). The project counted for 25\% of the overall course grade. 

Teams were given a score of up to 10 points: up to 1 point for a good VizItCard presentation (essentially demonstrating the consideration of alternative designs and specified goals); 2 points for functionality (how much of the solution worked); 3 points for design (overall execution of good visualization principles and usability); 1 point for interactivity features; 1 point for aesthetics; and 2 points for a good video presentation (with justification of choices, etc.). Teams winning the tournament or their round were given a small bonus (up to 5\% of the project or 1.25\% of the course). In the 2021 variant of the course, all teams received full points for the VizItCard activity and functionality (all projects worked). While most teams correctly implemented interactive features, roughly a quarter only received 50\% for the interactivity element. These teams either overly-emphasized static charts or did not provide useful interactions that furthered their design. A slightly smaller fraction (21\% of the class) received deductions (10\%-20\% off) for their design. This happened in cases where poor encodings were used and not argued for. A small subset (10\% of the class) only received 50\% for their aesthetic choices. In these cases the visualizations, while functional, appeared `rough' with vestigial elements. Most video reports received full scores. These distributions serve as valuable feedback to us, as instructors, in adding targeted training and activities for future years. We note that these results lend support to both individual students and programmatic achievement of our learning objectives for Roboviz (Section~\ref{sec:context}).

In addition to the rubric, we use an anonymous peer-assessment instrument. We use this for edge cases as well as for student recommendation letters. Each student is asked to allocated \$5000 bonus `dollars' among their team members (including themselves) and to explain their allocation.  Most teams produced a fairly equal distribution of dollars. Students in our program are familiar with collaborative work and know how to equitably divide and complete tasks. Our use of multiple game strategies also supported a more equal division of labor.

\section{Discussion}
In our analysis of student work we observe significant coverage of the topics covered in the course. That is, we see that students are successfully able to apply their learning in an integrated way. The final visual products (summarized in Figure~\ref{fig:types}) are `cross-cutting.' Roboviz allows students to implement different kinds of visualizations, interactions, and to consider a holistic `solution'. Our experience validates that both the intrinsic and extrinsic motivations in this project design help lead students to functional products that are well-motivated. The assessment of the value of any particular design decision is clear: does it help our team win or not?

Roboviz represents one possible type of game-centered project experience. There is certainly a cost to developing and debugging the infrastructure needed to run a project of this type. Thus, it would likely be difficult to create a new game every semester. Ideally, the game itself can be extended and enhanced in subtle ways. We attempted to do so with Roboviz by connecting strategies to different data types. In theory, we could add a stream of text (e.g., the robot's poetry can hint at their productivity) or geographical positions can be linked to the random-number generator for the guessing game. We encourage others to extend and remix the game design with `expansion packs' to offer additional variants. Additionally, we would like to see novel game concepts emerge with different strategic designs. Below, we reflect on some of the motivations for creating Roboviz, how the game can be used in the context of a course with diverse student interests, issues of strategy and complexity and, ultimately, a reflection on fun.

\subsection{Open-Ended Projects}
 
There is some benefit to instructors in using open-ended projects as it makes the initial steps of the project easier (i.e., we can provide very limited guidance). For instructors in research-focused settings, student projects can also occasionally lead to publications. However, the downside of this approach is that it becomes difficult to guide a large number of groups through the different challenges they encounter as the project progresses~\cite{burch2020more}. Perhaps more critically, it is difficult to create a well-developed rubric to assess student learning in the context of open-ended projects. Having run both types of projects in our course, we have found Roboviz to be the preferable option both for students and for the teaching staff.

That said, some students have a specific vision when taking the course. To support these students, we have allowed teams to propose alternatives. Only two groups--one team in 2020 and one individual in 2021--took us up on this offer. As long as handling the additional workload is possible, we believe this to be a reasonable accommodation.

\subsection{Strategies and Diversity}
Working in teams is challenging in the best of times. With the transition to online teaching, and with students spread throughout the world, we sought to further reduce the dependencies on team coordination. The initial version of Roboviz was designed so that different team members could tackle different sub-strategies, and thus visualizations. That is, one team member could work on network visualizations and another on hierarchical. These created `natural' partitions in the overall design. Teams only had to find a way of coordinating the bets and hacker requests and could otherwise function independently in both implementation and during game play. Because of the speed in which the game moves, there was some incentive to having multiple people monitoring different aspects of the dashboard.

While this encouraged a diversity of visualizations, the mechanics neither compelled nor particularly incentivized the creation of novel encodings. Thus, very few teams combined different data types into one visualization nor defined particularly novel interactions (beyond standard brush-over connectivity or linked filtering). 

One possible adaptation is to reward teams who come up with more novel encodings (e.g., through extra credit). Alternatively, one can compel the creation of combined visualizations by making it a project requirement. Currently, teams are told to cover at least two of the sub-strategies with the implication that these are done through two distinct visualizations. However, an instructor could specify that at least one the visualizations must combine two data types. The disadvantages of this approach are that (a) this may not be strategically beneficial (i.e., a combined encoding might be worse than two partitioned visualizations); and (b) the incentives for production of the solution become misaligned (i.e., the students are creating the solution because it's a requirement, not because it helps them play). Additional changes to the game mechanics might be needed to ensure that novel, combined encodings are both possible and interesting for the students to pursue.

\subsection{Implementation Challenges}
Live tournament play is fun but also presents some technical difficulties. Some issues also emerge due to misunderstanding of the game play. For technical difficulties, many teams had problems with connecting to a live remote server, rather than their local test environment. Depending on how quickly students can debug these issues, additional scaffolding such as a video walkthrough or pre-tournament test sessions or office hours are important. Another option we have considered is to allow the tournament to work over a longer period. This would allow students to update their strategies after testing them in actual game play. Unfortunately, this requires more time allocated for the project and would likely require removing other valuable assignments.

We found that there are a number of negative side effects to providing complex game rules. While the rules were explicitly designed to ensure different strategies--and therefore visualizations--we noticed that some students developed wrong models of the rules. For example, one team thought that the family tree showed robots with similar part structure (and therefore, productivity). This was not the case in our implementation. These misunderstandings provide an opportunity to add new features (e.g., robots from a given family \textit{should} have similar parts). However, they also point to the need for careful presentation of the rules and debugging of student understandings. One adjustment for the future is to have an early meeting with teams to check their understanding of the system and rules.

Finally, we found that some teams pushed the `no automation' rule. That is, teams were not supposed to \textit{automate} the game. They were supposed to create visualizations to help them make guesses, then manually put the guesses into the interface. However, many of our students are trained in statistics, data science, or machine learning. Their natural inclination is to model the data computationally rather than trying to build good visualizations. While emphasizing the no automation rule can be helpful, we have also begun to create more varied match configurations. These make it difficult to utilize one modelling strategy (e.g., the time series data doesn't fit a standard polynomial or high productivity robots can expire in a biased way--either late or early in the game). By varying the possible matches, teams may become more incentivized to make sure the human player is a more prominent decision maker.

As described earlier, we have elected to implement Roboviz as a real-time game as it simulates many analytical tasks where decisions and analyses vary over time. However, it should be possible to modify Roboviz to work on a data `snapshot' where the data is static. For example, teams can formulate their best bets and submit them within a certain time limit. This removes certain time pressures on a team but also removes some flexibility for the `game designer'. A temporally adaptive system may enable more variations as strategies need to be changed over time.

\subsection{On Fun and Games}
Whether gamified or game-centered, Roboviz was built on the idea that games can be motivational as they gratify self-determination or `Player Needs for Self Satisfaction (PENS)'~\cite{rigby2011glued}. In this model, games are found to be satisfying when they engage the sense of `competence, autonomy, and relatedness.' Roboviz evokes this by allowing teams to craft the tools necessary to achieve competence (simultaneously in the game as well as in the visualization discipline), autonomy (in the ability to control the outcome--teams are building for their own use), and relatedness (teams are collaborating and connecting both within and between teams). The idea that competitive or competitive/cooperative structures in motivating learning can be effective is certainly not new (e.g.,~\cite{slavin1996research}). That is not to say that that competitive structures are not problematic~\cite{kohn1992no}. Extrinsic and intrinsic motivations interact in potentially negative ways. Deadlines, competitions, directives, and so on can all reduce intrinsic motivation and thus impact learning~\cite{ryan2000intrinsic}. Clearly we cannot eliminate all directives, deadlines and even competition from a classroom. However, by disentangling successful game-play in Roboviz from the ultimate grade and encouraging effective intra-group cooperation~\cite{cohen1994restructuring} we hope to reduce these negative characteristics. Work with younger students also demonstrates that mixed cooperative-competitive games (e.g., the Teams-Games-Tournaments structure~\cite{devries1980teams}) are better than individually competitive or purely cooperative learning~\cite{okebukola1985relative}. We have worked to ensure the game is inclusive (e.g., by reducing the impact of game play on the grade). However, it is hard to conclude if we were successful given the relatively small population of students. We strongly believe in the importance of continued analysis of the inclusiveness of different project types.

As a final note on games, we acknowledge that it is difficult to create a game that provides `nudges' towards certain kinds of designs, provides an opportunity for actual assessment, and is simultaneously `fun.' These potential tradeoffs are familiar to anyone working in gamification or gamified education~\cite{burke2016gamify}. While we were inspired by traditional game design principles~\cite{schell2008art}, these were secondary to educational concerns. A hallmark of good game design is iteration through play-testing and updates. As such, future work can attempt to increase the quality of the game itself. This can include better themes, better visual elements (e.g., a `game board' that isn't just a score sheet), better balance between strategies, and additional variation that can enhance repeated game play. In our current implementation of Roboviz, luck has very little impact on the student's final grade. In some sense, the team's game strategy does not necessarily have a huge impact on the grade either. We view this as a positive as it incentivizes good design above all else. However, these aspects (e.g., luck, strategy) can be further emphasized or de-emphasized through other mechanics. For example, having more of the grade based on game performance or using round-robin style playoffs instead of tournament. In part, our goal was to make Roboviz flexible to these other objectives. We hope that others consider the use of game-based projects for visualization courses and develop alternative games.

\subsection{Context and Limitations}
One of the benefits of our broader program is that many of the topics that we consider visualization adjacent are covered in other courses. For example, data cleaning and client engagement are covered in specific courses that are largely prerequisites to our class. Programs that do not have this type of overall curricular structure may have different learning objectives than ours (e.g., students will be able to clean data for use in visualization systems; or students will be able to scope a project with their clients). It may be possible to modify the game structure of this course to emphasize these other learning objectives. For example, one could introduce noisy data or poorly document APIs. To simulate a client-driven experience common in many HCI courses (e.g.,~\cite{koppelman2006creating}) one could recruit a set of game players external to the course who would need to use the interface to play. 

While these modifications are possible we acknowledge that a more traditional open-ended projects may be more suitable here. With Roboviz, our goal has been to emphasize the design and implementation of effective and usable visualization systems. This comes with some cost to `realism' but provides an alternative strategy that focuses on a subset of infovis concepts. 

Our course also has the benefit of running over a 14-15-week semester. Students have already been instructed in how to implement visualization systems and the theory, data type, and encodings that are useful for the project have been covered before the students being building. For example, we cover geovisualization and text visualization in the last few weeks of the course--data types that are not part of the Roboviz game (network/hierarchical data is covered just as they receive the assignment). Shorter courses or ones that do not cover set of topics, in the same order, may benefit from modifying the mechanics of Roboviz and including a different set of data.

As far as we are aware, Roboviz is the first project of this type. Our hope is that other variations may develop so that the visualization pedagogy community can identify a broad set of guidelines (e.g.,~\cite{chatzis}) for the use and implementation of game-based projects.

\section{Conclusion}
In this paper we reflect on our creation and use of Roboviz, a game-based project for information visualization courses. Roboviz was designed to address limitations in traditional projects. The game was designed to (a) deliver clean data with clear goals, (b) challenge and motivate students (intrinsically and extrinsically) to use their learning and skills, and (c) provides a clear assessment framework for both students and course staff. Our experience with the project over two semesters has been a very positive one. Pedagogically, we believe projects like Roboviz has numerous advantages and hope to see other instructors use the materials or develop their own game-centered projects.
\acknowledgments{We would like to thank our students for (a) being patient participants in the first runs of Roboviz (which occasionally required quick patches and all in the context of COVID-related shutdowns), and (b) allowing us to share their work in this paper. We'd also like to thank Licia He for helping run the Fall 2020 version of the course, and Barry Fishman, Ben Shneiderman, and Mark Guzdial for helpful pointers and discussion. Finally, we'd like to thank our reviewers for their challenging and useful feedback. This work was partially funded by NSF IIS-1815760.}

% BALANCE COLUMNS
\balance{}

% REFERENCES FORMAT
% References must be the same font size as other body text.
\bibliographystyle{abbrv-doi}
\bibliography{08_biblio}

\end{document}